\documentclass[twocolumn]{aastex631}

\usepackage{CJK}
\usepackage{amsmath}
\usepackage[multiple]{footmisc}

\def\sixteen{\text{\rm AGNUV4}}
\def\twelve{\text{\rm AGNUV0.03}}
\def\sixteenB{\text{\rm AGNUVB0.6}}
\def\eighteenB{\text{\rm AGNUVB3}}

\def\msun{M_\odot}

\shorttitle{Magnetosonic Waves in AGN Disks}
\shortauthors{Kaul et al.}

\begin{document}
\begin{CJK*}{UTF8}{gbsn}

\title{Magnetosonic 
Waves as a Driver of
Observed Temperature Fluctuation Patterns\\ in AGN Accretion Disks}

\author[0009-0001-1399-2622]{Ish Kaul}
\affiliation{Department of Physics, University of California, Santa Barbara, CA 93106, USA}

\author[0000-0002-8082-4573]{Omer Blaes}
\affiliation{Department of Physics, University of California, Santa Barbara, CA 93106, USA}

\author[0000-0002-2624-3399]{Yan-Fei Jiang (姜燕飞)}
\affiliation{Center for Computational Astrophysics, Flatiron Institute, New York, NY 10010, USA}

\author[0000-0003-0232-0879]{Lizhong Zhang (张力中)}
\affiliation{Center for Computational Astrophysics, Flatiron Institute, New York, NY 10010, USA}



\begin{abstract}
Recent observations have revealed slow, coherent temperature fluctuations in AGN disks that propagate both inward and outward at velocities of $\sim 0.01-0.1 c$, a kind of variability that is distinct from 
reverberation (mediated by the reprocessing of light) 
between different regions of the disk. 
We investigate the origin and nature of these fluctuations using global 3D radiation-magnetohydrodynamic 
simulations of radiation 
and magnetic pressure-dominated AGN accretion disks. Disks with a significant turbulent Maxwell stress component exhibit 
wave-like temperature perturbations, most evident close to the midplane, whose propagation speeds 
exactly match the local fast magnetosonic speed and are consistent with the speeds inferred in observations. 
These fluctuations have amplitudes of $2-4\%$ in gas temperature, which are also consistent with observational constraints.  Disks that are dominated by mean-field Maxwell stresses do not exhibit such waves.  While waves may be present in the body of the disk, we do not find them to be present in the photosphere.  Although this may in part be due to low numerical resolution in the photosphere region, we discuss the physical challenges that must be overcome for the waves to manifest there.  In particular, the fact that such waves are observed implies that the disk photospheres must be magnetically dominated, since radiative damping from photon diffusion smooths out radiation pressure fluctuations. Furthermore, the gas and radiation fluctuations must be out of local thermodynamic equilibrium.  
\end{abstract}



\section{Introduction} \label{sec:intro}
Active Galactic Nuclei (AGNs) are powered by accretion of matter onto supermassive black holes in the centers of galaxies, 
and play a crucial role in regulating galactic evolution through energetic feedback processes \citep{fabian2012, kormendy2013, harrison2018}.
The nature of this feedback depends on the physics of the accretion flow, for which there are currently many uncertainties. Optical-near infrared spectra \citep{kishimoto2008}, reverberation mapping \citep{cackett2021}, and microlensing studies \citep{morgan2010} are qualitatively consistent with a radial stratification in temperature with thermalization of accretion power in an optically thick medium resembling an accretion disk.  However, quantitative size scales are generally a bit larger than predicted by standard accretion disk theory.  Moreover, the energetically important far ultraviolet to soft X-ray spectral energy distributions are generally inconsistent with standard accretion disk theory \citep{davis2007,kubota2018}.

One of the primary ways to study AGN 
is by analyzing the temporal continuum variability in X-ray, UV and optical bands. Such variability offers key insights into the structure, dynamics and physical processes of the accretion flow. 
While extensive observational studies \citep{geha2003,macleod2010,burke2021} and recent simulations \citep{davis2020} have provided valuable constraints and insights, the precise origin of this variability remains an open problem.

One widely used approach to studying AGN variability is reverberation mapping, first proposed by \citet{blandford1982} in the context of measuring light travel times in and between 
the disk and the broad line region. 
This technique leverages time lags between different wavelength bands to infer the radial profile of 
temperature in the photosphere of the accretion flow 
\citep{sergeev2005,shappee2014, edelson2015, edelson2017}. This method assumes that X-rays, which are emitted from a compact corona close to the black hole, are reprocessed by the disk, which drives temperature fluctuations that respond on the light crossing time from the corona to the disk \citep{krolik1991, haardt1991, frank2002}.  
Thus, correlated behavior should be observed across different wavelengths, but with a lag that increases with wavelength, reflecting the cooler emission from regions further out in the flow.  
Previous work has indeed measured 
this lag behavior in 
the optical and UV 
continuum \citep{sergeev2005, edelson2015, edelson2017, mchardy2014}. 

However, numerous studies have shown that the behavior is not always so simple.  
For instance, while the X-ray emitting corona is presumed to be the originating source of variability, 
X-rays can lag optical emission or just show no signs of correlation \citep{kazanas2001, edelson2019, dexter2019, cackett2020}, bringing into question the idea that the reprocessing of X-rays 
is the only way optical/ultraviolet variability manifests. \citet{dexter2011} proposed that inhomogeneous temperature fluctuations in the disk naturally result in correlated variability across different wavelengths on short timescales (i.e. shorter than the viscous time), although they did not propose a concrete physical mechanism for generating such fluctuations and sustaining them against orbital shear. This resolves the inferred optical disk size discrepancy from reverberation 
and microlensing studies and more standard disk theory. 
Flux variations could also be caused by changes in stresses 
in the disk which result in accretion variability \citep{lyubarskii1997, arevelo2006}. However, the inflow (viscous) timescale from standard accretion disk theory \citep{shakura1973} is too long, $\sim 100$ years \citep{davis2020}, compared to observations that see short timescale variability. Opacity-driven convection could also be responsible for flux variability due to the enhancement of opacity in the UV/optical bands driving changes in turbulent stresses on the thermal timescale $\sim100$ days \citep{jiang2020}. 
Thermal fluctuations driven by other intrinsic processes 
have also been proposed to explain variability \citep{cai2018,sun2020}. Furthermore, these fluctuations have been modeled using damped random walks \citep{macleod2010, stone2022, burke2021, suberlak2021}, which correspond to thermal timescales and are correlated with the black hole masses. 
Thus, in addition to reprocessing by the disk, there are multiple possible sources of variability on timescales longer than the light-crossing time.

Recently, a new type of variability has been uncovered in AGN disks. \citet{neustadt2022}, for the first time, used time-series observations of UV/optical light curves to directly map the temperature structure and perturbations in AGN accretion disks. Their method assumes that the disk follows a standard thin-disk model \citep{shakura1973}, the disk emission is axisymmetric, and the temperature fluctuations are linear. Using these assumptions, they modeled light curves from seven AGNs using AGN STORM (Space Telescope and Optical Reverberation Mapping) data \citep{starkey2017}. They identified the existence of coherent temperature fluctuations that correspond to both outgoing \textit{and ingoing} waves that move at velocities of $\sim 0.01 c$, ruling out mere reprocessing on the light crossing time as an explanation. 
These fluctuations also occur over relatively short timescales of $\sim$few - 100 days, which makes them 
easily observable. This methodology was then used by \citet{stone2023} on SDSS-RM (Sloan Digital Sky Survey - Reverberation Mapping) \citep{shen2015, shen2019} quasar data to again confirm the existence of coherent slow-moving temperature fluctuations in luminous intermediate-redshift ($z<4.5$) quasars. The same result was then seen and extended for a Seyfert 1 galaxy \citep{neustadt2024}. 

With the advent of fast and accurate computational methods for solving magnetohydrodynamics (MHD) problems coupled with detailed radiative transfer, it is now possible to simulate AGN accretion flows 
across multiple scales to make predictions about the accretion flow 
structure, its dynamical evolution and its fluid as well as radiative properties \citep{jiang2019, jiang2020, guo2023, hopkins2024a, jiang2025}. Such global AGN studies have been instrumental in probing the large scale structure of disks, outflows such as jets and magnetocentrifugal winds, angular momentum transport models including magnetorotational (MRI) 
turbulence, and spectral formation. 
Recent local shearing box simulations have shown that the disk turbulence, instead of hard X-rays with scattering dominated opacity, is sufficient 
to produce the observed variability in reverbation mapping data \citep{secunda2024}. These simulations with multi-frequency opacities also examine and compare the X-ray-UV lags to data \citep{secunda2025}. It is thus vital to explore the assumptions and results of reverberation mapping campaigns with simulations to infer the underlying physical processes.
Here we present the results of global radiation MHD simulations of AGN accretion flows, 
describing their thermal properties and elucidating the nature of wave-like thermal fluctuations to compare with the recent slow-moving temperature perturbations that have been observed 
in UV/optical bands \citep{neustadt2022,stone2023,neustadt2024}.

The remainder of the paper is organized in the following manner. Section \ref{sec:methods} presents the simulation setup along with codes and methods used to perform the simulations. Section \ref{sec:results} shows the results of the simulations, including a discussion of 
the overall structure and evolution, as well as an 
analysis of the origin and properties of temperature fluctuations. Finally, Section \ref{sec:discussion} demonstrates how this variability might 
distinguish thermal pressure-dominated and magnetic pressure-dominated photospheres of AGN disks with challenges in observing them at the photosphere, as well as how more simulation runs covering a wider parameter space can improve upon this distinction and their observed features. 

\section{Overview of Simulations} 
\label{sec:methods}

We discuss five simulations in this paper, all run using the radiation MHD code \textsc{Athena++} \citep{stone2008}.  These are a radiation pressure dominated simulation that exhibits strong iron opacity-driven convection (AGNIron, \citealt{jiang2020}), and four new simulations: \eighteenB, \sixteenB, \sixteen, and \twelve\ \citep{jiang2025}. The setup of the simulations is similar to that described in \citet{jiang2019}, and more details can be found in \citet{jiang2020} and \citet{jiang2025}.  All the simulations use a spherical polar coordinate $(r,\theta,\phi)$ grid.  Simulation AGNIron uses a black hole mass of $M=5\times10^8\msun$, a simulation time unit of $t_{\rm sim}=10^7$~s, and an inflow inner boundary condition at $r=30r_{\rm g}$, where $r_{\rm g}=GM/c^2$ is the gravitational radius of the hole.  The other simulations use $M=10^8\msun$, a simulation time unit of $t_{\rm sim}=4\times10^6$~s, and an inflow inner boundary condition at $r=50r_{\rm g}$.

The simulations are initialized with a weakly magnetized torus: simulations \eighteenB\ and \sixteenB\ are initialized with two poloidal loops of magnetic field above and below the midplane, while the other three simulations are initialized with a single poloidal loop.  For reasons related to this initial magnetic topology and the initial density and entropy within each torus, simulations AGNIron and \eighteenB\ accrete inward to produce a strongly radiation pressure dominated (i.e. high plasma beta) accretion flow in the midplane regions.  Simulations \sixteenB\ and \twelve\ evolve into magnetically elevated flows with low plasma beta in the disk midplane regions.  Simulation \sixteen\ accretes inward to form a radiation pressure dominated flow to begin with, but strong angular momentum losses in a magnetocentrifugal wind drive rapid accretion, lowering the surface mass density and producing a magnetically elevated flow. 

Many more details of the physical properties of these five simulations can be found in \citet{jiang2020} and \citet{jiang2025}.  From now on in this paper, we focus on those properties most relevant to the presence or absence of acoustic waves.

\begin{figure*}
    \centering
    \includegraphics[width=\linewidth]{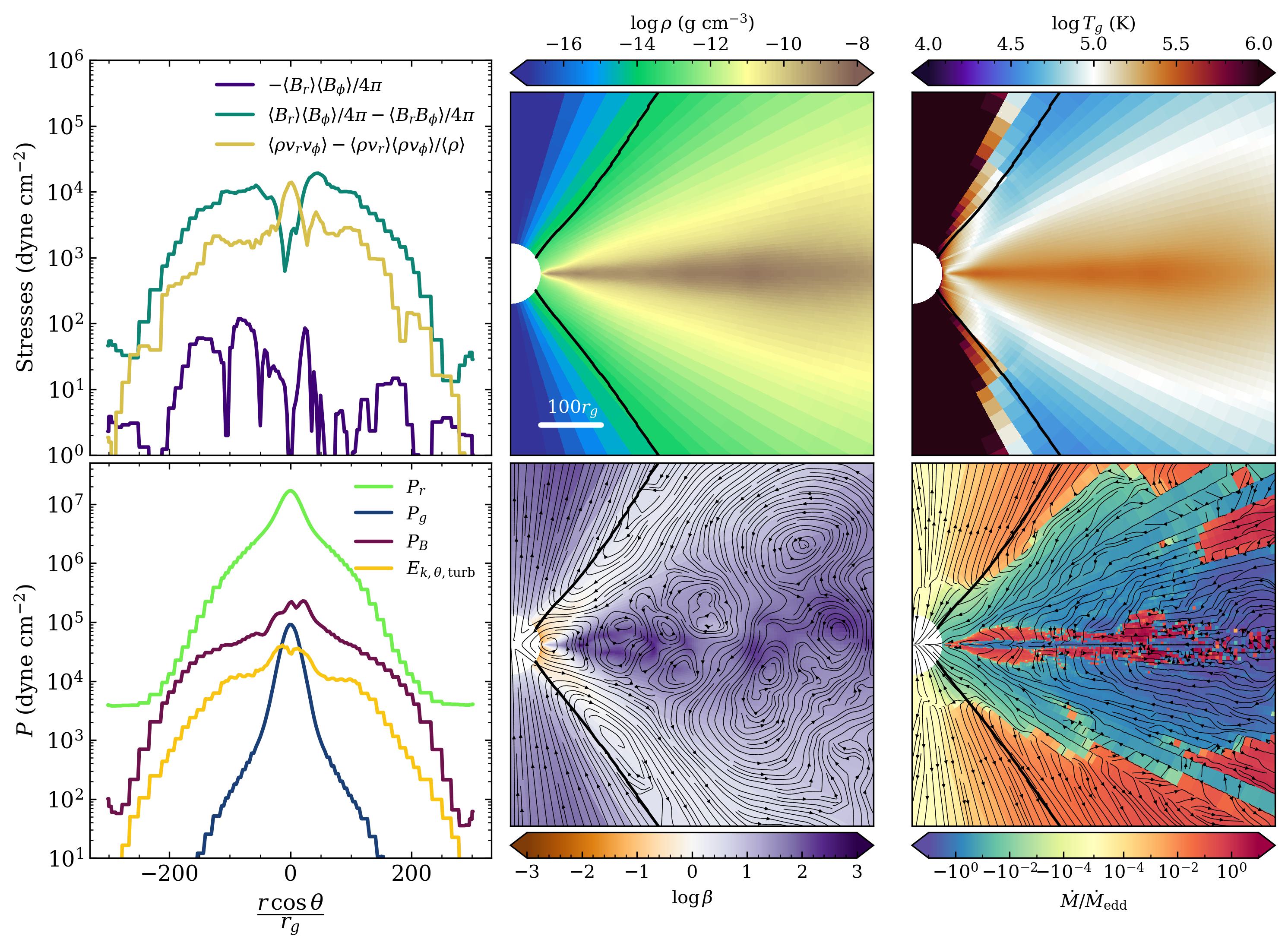}
    \caption{Azimuthally- and time-averaged quantities after the initial evolution of the \eighteenB\ simulation. (Left Top) Mean ($-\langle B_r\rangle\langle B_{\phi}\rangle/4\pi$) and turbulent ($\langle B_r\rangle\langle B_{\phi}\rangle/4\pi - \langle B_r B_{\phi}\rangle/4\pi$) Maxwell stress along with turbulent Reynolds ($\langle \rho v_r v_{\phi}\rangle - \langle\rho v_r\rangle\langle\rho v_{\phi}\rangle/\langle\rho\rangle$) stresses and (Left Bottom) pressure profiles in $z$ at $r=300r_g$ averaged over $100-200t_{\rm sim}$. The pressure profiles include radiation, gas and magnetic pressure in addition to the $\theta$ component of the turbulent kinetic energy. The disk is dominated by radiation pressure everywhere, although magnetic pressure and turbulent kinetic energy become substantial farther away from the midplane. Other panels show the 
    distributions of density (Middle Top), gas temperature (Right Top), plasma $\beta$ overplotted with poloidal magnetic field ($B_r, B_\theta$) streamlines (Middle Bottom), and mass accretion rate in terms of the Eddington rate overplotted with the poloidal velocity ($v_r, v_\theta$) streamlines (Right Bottom). The thick black lines in the 2D maps are the northern and southern photospheres, defined as the regions where the optical depth, computed from the Rosseland mean opacity, is unity. The carved out white semi-circle denotes the $r=50r_{\rm g}$ inner boundary, and its center is the origin where the black hole is placed. The disk is supported 
    by radiation pressure and the magnetic field line eddies in the disk are indicative of high $\beta$ MRI turbulence. The region around and beyond the photosphere has evidence for a wind due to the positive mass outflow, whereas the rest of the disk is dominated by accretion flow into the black hole.}
    \label{fig:theta_profile}
\end{figure*}
\section{Results}
\label{sec:results}

\subsection{\eighteenB\ Overall Disk Structure}

As we will describe below, the presence of acoustic waves is most evident in the two thermal pressure dominated disks AGNIron and \eighteenB.
We therefore begin by discussing 
the overall structure of the latter here, treating it as our fiducial simulation.  
The left panels of \autoref{fig:theta_profile} illustrate the azimuthally-averaged 
Maxwell and Reynolds stresses
along with (various) pressure profiles as a function of vertical height ($z$) at a fixed radius ($r=300 r_g$), averaged over a $100 t_{\rm sim}$ time window. The turbulent and mean Maxwell stresses are plotted separately. We also plot the $\theta$ component of the turbulent kinetic energy which we calculate as $E_{\rm k,  \theta, turb}=\langle\rho v_{\theta}^2 - \overline{\rho v_{\theta}^2}\rangle$, where $\langle\cdots\rangle$ represent average over azimuthal angle $\phi$ and time $t$. These images correspond to times 
after the MRI has fully developed. Note that inflow equilibrium is only established inside 200$r_{\rm g}$ in this simulation (see Figure 2 of \citealt{jiang2025}).  Nevertheless, the fluctuations that we identify below propagate at speeds much faster than the inflow speed, and have intrinsic periods that are also much shorter than the inflow time.  They are therefore unlikely to be affected by the lack of inflow equilibrium of the disk.

The disk is highly turbulent, and the primary mode of angular momentum transport is by turbulent Maxwell stresses, with a lower but significant component from the turbulent Reynolds stress. In the very central densest region of the midplane, this dominates over the Maxwell stress. 
The disk is dominated by radiation pressure across all vertical layers, although magnetic pressure and turbulent kinetic energy increase in importance at approximately $200 r_{\rm g}$ above and below the midplane. The middle and right panels in \autoref{fig:theta_profile} show the poloidal, azimuthally-averaged maps of various quantities of the disk. The density is concentrated at the midplane and falls off away from it, although the photospheres are much further away from the midplane. The disk is 
puffed up by radiation pressure, as evidenced by the plasma $\beta$ map which shows that the disk is mostly weakly magnetized except for regions very close to the black hole and for the regions close to the photosphere (which still are at about $\beta\sim 1$ and never lower). 
The magnetic field lines (overplotted on the $\beta$ map) show eddies inside the disk close to the midplane, which is evidence for MRI turbulence. 
The toroidal magnetic field shows flipping in polarity in time, consistent with weak-field MRI butterfly diagrams 
\citep{brandenburg1995}. 
Furthermore, mass is accreting into the black hole in most regions of 
the disk (at 
Eddington or super-Eddington rates in some regions near 
the midplane) except close to and beyond the photosphere, where there is a weak outflow that traces the magnetic 
field lines. 
However, close to the midplane there is a rapid flipping of polarity of the mass inflow rate, which is in fact intricately tied to temperature fluctuation waves as we shall examine in the next section.

\subsection{Temperature fluctuations}

\begin{figure}
    \centering
    \includegraphics[width=\linewidth]{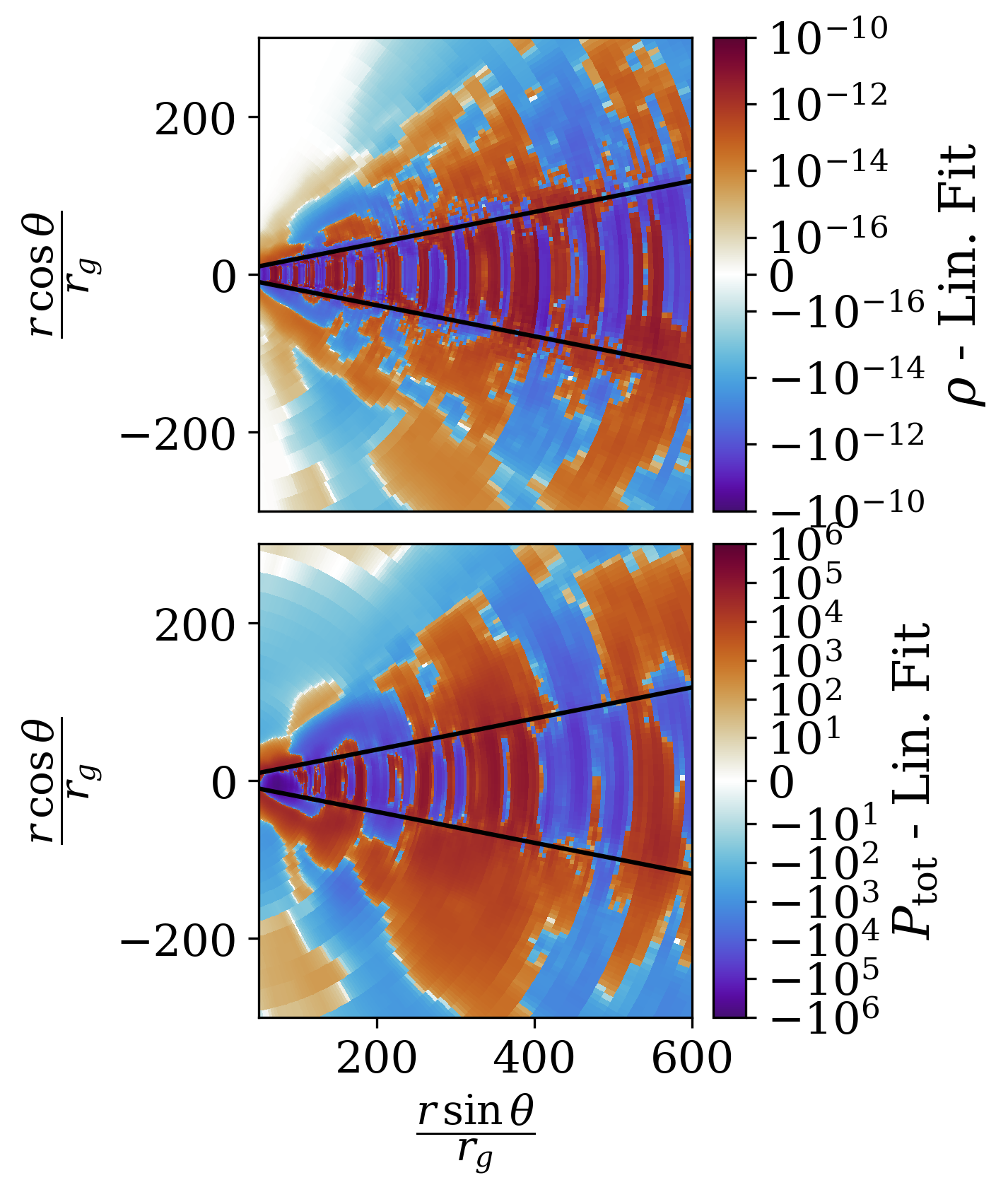}
    \caption{Snapshots of azimuthally-averaged data from the simulation \eighteenB\ at $225 t_{\rm sim}$, showing evidence of waves near the midplane.   The top panel displays the density with a temporal linear fit over a window of $50 t_{\rm sim}$ subtracted. 
    The bottom panel illustrates that these waves also appear in the total pressure (magnetic, gas, and radiation) fluctuations defined in the same way. 
    Black lines in both panels show the numerical resolution refinement boundary between the second and third refinement levels. The wave patterns appear to be confined to within this refinement boundary, suggesting that high spatial resolution is needed to observe them.
    A video of the time evolution of these fluctuations can be viewed \href{https://doi.org/10.5281/zenodo.15053915}{online: doi.org/10.5281/zenodo.15053915}. A similar video for \sixteenB\ can also be viewed \href{https://doi.org/10.5281/zenodo.15498854}{online: doi.org/10.5281/zenodo.15498854}}
    \label{fig:wave_zoom}
\end{figure}

We now 
examine the disk for evidence of temperature fluctuations and their nature. Snapshots of azimuthally-averaged density and total pressure fluctuations, as shown in \autoref{fig:wave_zoom}, indicate the presence of wave-like patterns near the midplane.   These plots show $\langle P_{\rm tot}\rangle_{\phi}$ (the azimuthally averaged total pressure) minus $P_{\rm lin, 50 t_{\rm sim}}$, which 
is a linear fit in time to $\langle P_{\rm tot}\rangle_{\phi}$ 
within a $50 t_{\rm sim}$ window centered on the snapshot at 225 $t_{\rm sim}$. The same procedure was also done for the density map. 
This temporal linear fit subtraction was done in order to reveal 
these waves more clearly.\footnote{Note that not all visible fluctuations in this figure correspond to propagating waves. Some of the fluctuations are static and have propagating patterns that move over them. True propagation signatures can be found in the time-stream plots below.} 
These fluctuations propagate predominantly due to thermal pressure variations, 
as we discuss 
below.

\begin{figure*}
    \centering
    \includegraphics[width=\linewidth]{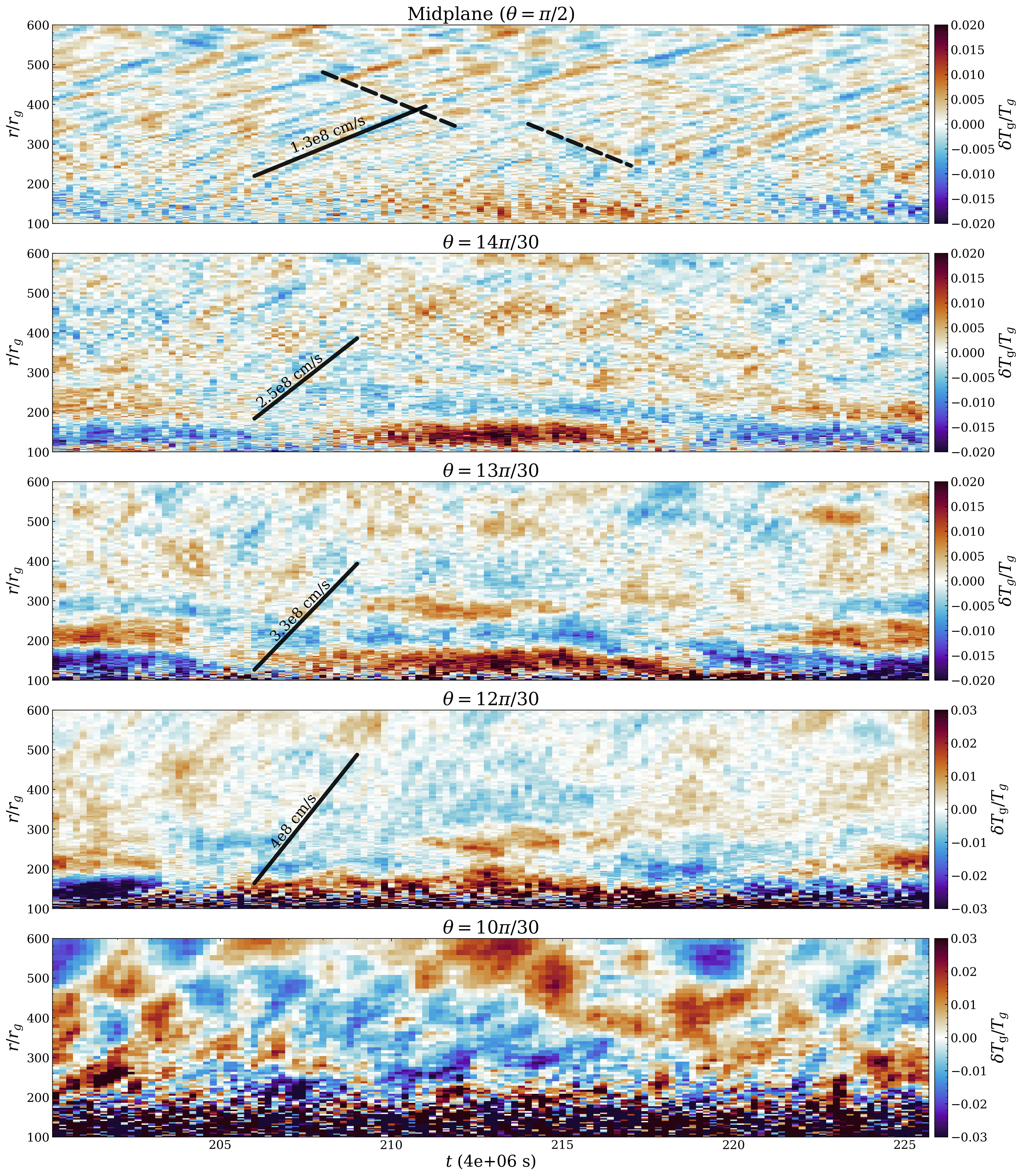}
    \caption{Timestream ($r-t$) plots of gas temperature fluctuations $\delta T_g$ over the mean $T_g$ in simulation \eighteenB, for the time interval shown and for selected values of $\theta$ from the vertical as labeled in each panel. The over-plotted black lines indicate the magnetosonic sound speed $c_{\rm ms}$ at those radii, showing that these are acoustic waves. The dashed lines in the midplane guide the eye towards ingoing fluctuations. 
    The coherent fluctuations associated with these waves do not appear to be present in the bottom panel ($\theta=\pi/3$), indicating that they are confined to the midplane regions.}    \label{fig:timestream_theta}
\end{figure*}

\autoref{fig:timestream_theta} presents 
(radius - time) timestream plots of relative fluctuations in azimuthally-averaged gas temperature
$\delta T_g/ T_g$ for selected polar 
angles $\theta$. These are taken between a time window of 200 and 226 $t_{\rm sim}$, subtracting out the mean from this window.  Since all our wave frequencies are much higher than the orbital and epicyclic frequencies, we can treat the disk as a static medium and use time averaged data at late times. We shall make use of this assumption in the analysis below. These wave signatures can now clearly be seen as fluctuations in gas temperature propagating outwards (and inwards in the midplane plot) with a decreasing slope (i.e. slower speed) further out in radius.  The amplitudes of the fluctuations, are around $~2-4\%$ above and below the mean, and are consistent with observations of AGN 
\citep{neustadt2022, neustadt2024, stone2023}.

The black lines in \autoref{fig:timestream_theta} correspond to propagation at the local magnetosonic speed ($c_{\rm ms} = \sqrt{c_{\rm s}^2+v_{\rm A}^2}$). 
In this simulation, 
the dominant form of pressure in 
this speed is radiation pressure. The agreement between the slopes of these curves and the outward propagating temperature fluctuations indicates 
that the 
wave signatures originate from thermal pressure fluctuations in acoustic waves. 
These waves also demonstrate a coherence on the order of $\sim 10$s of days, which again, is well within the bounds provided by observations \citep{neustadt2022}. While an FFT would be a natural choice to isolate this coherence period, we find that both 2D and 1D FFTs do not clearly show any dominant modes that can be pinpointed as the wave occurrence frequencies. This suggests that that there are other fluctuations that co-exist and drown out the actual thermal fluctuation wave signal or the waves are not coherent with a single occurrence frequency.  

\autoref{fig:timestream_theta} also shows that as the angle from the midplane is increased (decreasing $\theta$ in the northern hemisphere shown), the slope of the black line also increases, indicating higher $c_{\rm ms}$ at higher altitude to the point where closer to the photosphere the speeds come within the observed range of velocities ($\sim 0.01-0.1c$). However, we also find that waves are more prominent near the midplane and are non-existent at larger vertical angles.
This may be because we cannot fully resolve these waves up into the photosphere because it is too far out (see \autoref{fig:theta_profile} to see the puffed out structure of the disk) and has poor spatial resolution in our simulations (see the refinement boundary in \autoref{fig:wave_zoom}).  Another possibility is that there is a physical damping mechanism at work which prevents these waves from extending into the photosphere, and which would therefore prevent them from being observable.  We shall examine this below in the context of constant height slices.  

\begin{figure*}
    \centering
    \includegraphics[width=\linewidth]{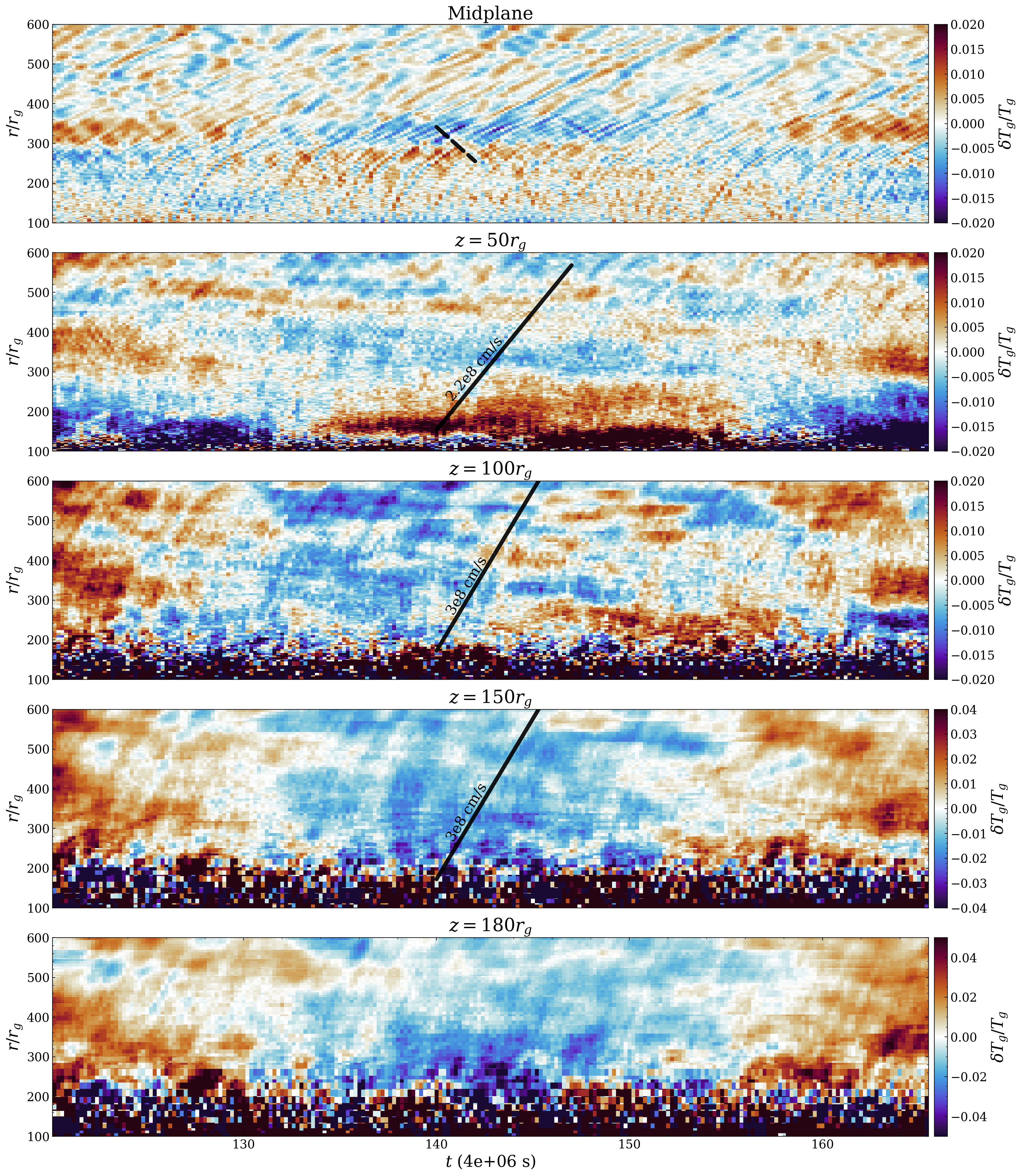}
    \caption{Same as \autoref{fig:timestream_theta} but for slices in constant vertical height $z$. The midplane shows both ingoing and outgoing wave signatures. At $z>180r_{g}$ the signatures of these propagating fluctuations vanish. 
    }
    \label{fig:timestream_z}
\end{figure*}

\autoref{fig:timestream_z} shows an alternative analysis in which we examine 
slices at constant vertical heights $z$. The waves appear to be more easily identified here. This, however, is expected because these waves appear to have nearly vertical wavefronts 
in \autoref{fig:wave_zoom} which would make these oscillations more evident in slices in height as opposed to $\theta$.  In the constant $\theta$ case, we see the projection of the wave vector onto the $\theta$ slices.  Near the midplane, the simulation captures both ingoing and outgoing wave patterns. However, at higher vertical positions ($\sim 200 r_g$), wave signatures are largely absent in the inner radii.  This is at least partly because the waves are confined to a wedge (\autoref{fig:wave_zoom}), possibly because of the constant $\theta$ refinement boundaries.  Moreover, for higher $z$ slices, we should not consider the timestream at small radii, since this region is well outside the photosphere and away from the disk, where there is not much mass.
Similar to \autoref{fig:timestream_theta}, the overplotted black lines show the consistency of these waves being driven by fluctuations which propagate at the local magnetosonic sound speed. Here, in contrast to \autoref{fig:timestream_theta}, $c_{\rm ms}$ shows less variation for higher $z$ slices, which is consistent with the wavefronts being vertical in \autoref{fig:wave_zoom}.

To examine if radiation damping is responsible for the disappearance of these waves at large altitude as well as for small radii at altitude ($r<200r_g$ at $z=180r_g$ for instance), we plot the ratio of the radiation damping timescale to the wave propagation timescale
at the chosen height ($z$) slices in \autoref{fig:damping}. We compute the damping timescale from \autoref{eq:fastdamping} in the Appendix as $t_{\rm damp}=(1+v_{\rm A}^2/c_{\rm s}^2)3\kappa_{\rm R}\rho\lambda^2/(2\pi^2c) $, where $\kappa_{\rm R}$ is the Rosseland mean opacity, and $\lambda$ is the wavelength of the wave at that specific height. Similarly, the wave period
is $t_{\rm wave}=\lambda/ c_{\rm s}$ where $c_{\rm s}$ is the thermal sound speed. \autoref{fig:damping} shows two important features. First, at higher altitude $t_{\rm damp}/t_{\rm wave}<1$ for small radii. This implies that radiation damping becomes significant in these regions and therefore destroys 
wave signatures, a feature that is indeed observed in \autoref{fig:timestream_z}. Second, this ratio also has an overall decrease for higher $z$, implying that acoustic waves would become damped \textit{ in general} at altitude. We can thus conclude that propagating fluctuations driven by acoustic waves are unlikely to be observable at the photosphere if thermal pressure dominates there.

\begin{figure}
    \centering
    \includegraphics[width=\linewidth]{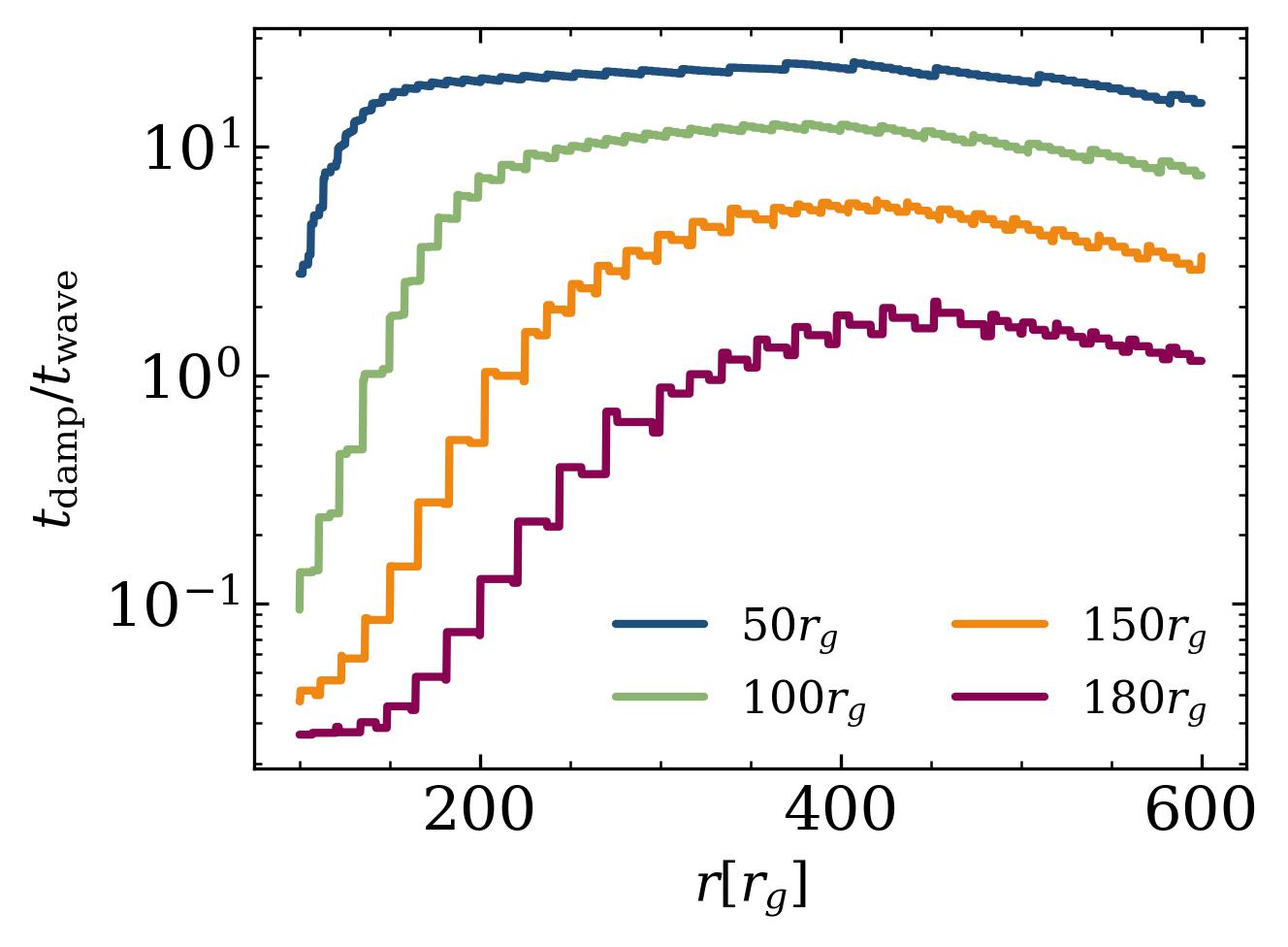}
    \caption{The ratio of the radiation damping timescale to the wave propagation timescale in \eighteenB\ at different height ($z$) slices. For higher $z$ this ratio goes below unity for small radii which causes the disappearance of these waves in those regions. Additionally, for higher altitude this ratio drops lower and lower.}
    \label{fig:damping}
\end{figure}

\subsection{Other simulation runs}

\begin{figure}
    \centering
    \includegraphics[width=\linewidth]{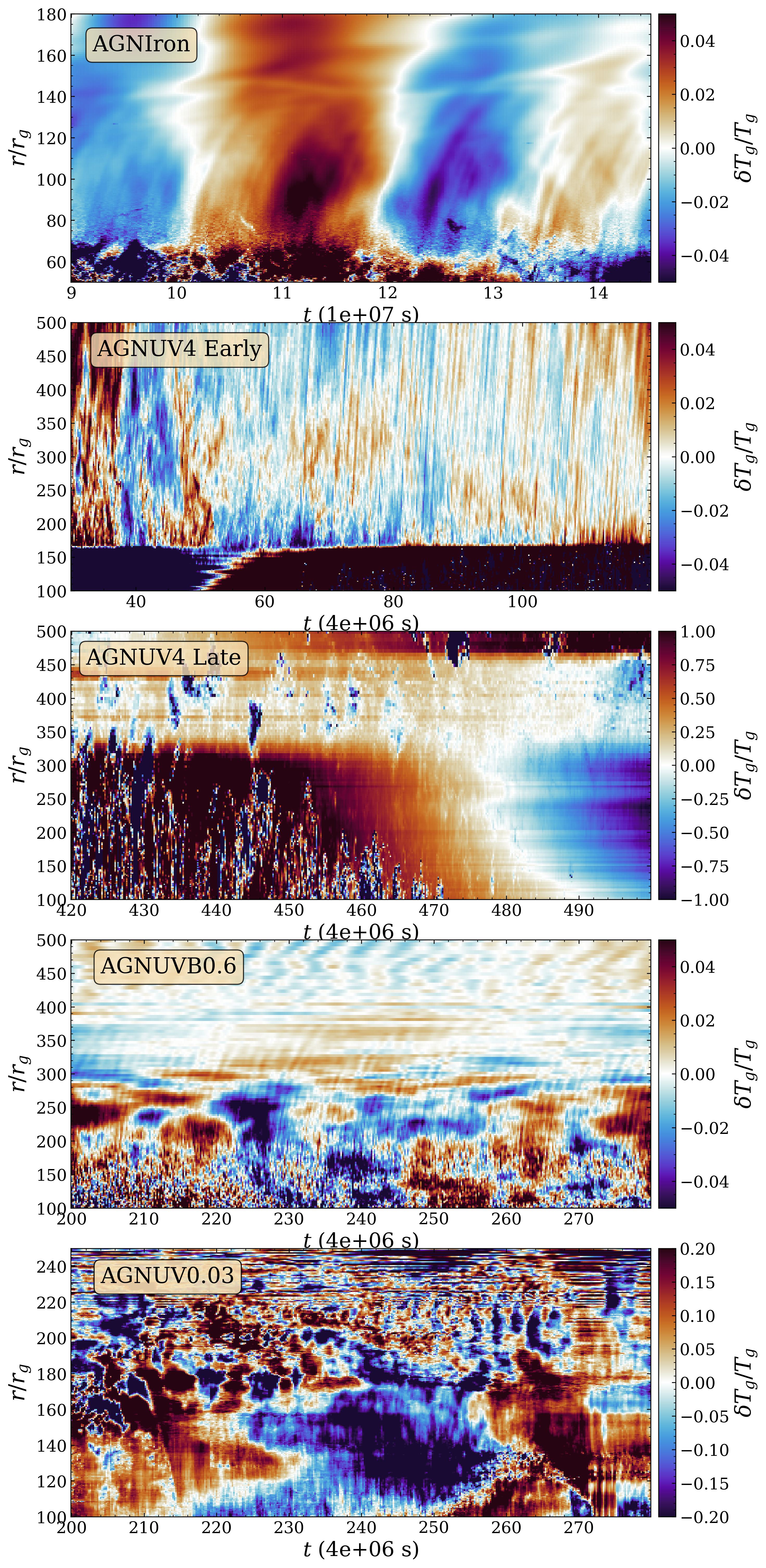}
    \caption{Timestream plots of midplane gas temperature fluctuations for four different simulations: AGNIron \citep{jiang2020}, \sixteen\ for early and late times, \sixteenB, and \twelve.  Each plot is detrended using a linear fit in time at each radius to highlight wave-like structures. AGNIron and \sixteen, which are thermal pressure-dominated (early time for \sixteen, after which it becomes magnetic pressure dominated in the midplane for larger and larger radii), exhibit clear propagating wave signatures. \sixteenB\ is magnetic pressure-dominated in the midplane up to $\sim 250r_g$, beyond which a thin thermal pressure-supported region emerges, along with the appearance of wave-like features
    above $200r_g$. \twelve, a magnetically dominated disk, lacks any significant wave structure. Both its fluctuations and those of \sixteen\/ at late times, where there is also no coherent wave propagation, are much stronger, as shown by the colorbar range increase.}
    \label{fig:timestream_compare}
\end{figure}

To assess whether acoustic waves are generally present in AGN accretion flows under different physical consitions, 
we examine additional simulation runs with varying pressure support conditions. \autoref{fig:timestream_compare} presents radius-time plots of gas temperature fluctuations for four distinct simulations: AGNIron \citep{jiang2020}, \sixteen, \sixteenB, and \twelve. These simulations explore different regimes of thermal and magnetic pressure dominance to determine the conditions necessary for wave-like temperature variations.

Like \eighteenB, AGNIron is a thermal pressure-dominated disk, but it is strongly influenced by the iron opacity bump, which induces repeating cycles of convection that cause changes in the MRI turbulent stresses \citep{jiang2020}.  
This cyclical process results in strong variability 
and surface density fluctuations on the local thermal time scale. Despite these differences with \eighteenB, the top panel of \autoref{fig:timestream_compare} shows clear evidence of propagating temperature fluctuations in the $\sim70-140r_g$ radial range at the midplane, with propagation speeds 
at the local sound speed 
$\sim 0.1c$ here.

\sixteen\ initially exhibits thermal pressure dominance in the midplane but transitions to magnetic pressure dominance over time. The inner boundary of the thermal pressure-supported region gradually moves outward as the disk surface density declines.  
The second panel of \autoref{fig:timestream_compare} shows the timestream for the early phase in this simulation, where thermal pressure still dominates. Similar to AGNIron and \eighteenB\/, this regime also produces propagating temperature fluctuations, but the later, magnetically elevated region (third panel) does not.  As we discuss further below, this is because the magnetically elevated region is dominated by mean field, rather than turbulent, Maxwell stresses which are unable to excite waves.  

\sixteenB\ produces a disk that is 
magnetically dominated everywhere 
except for a geometrically thin region near the midplane 
appearing only at radii beyond $\sim 200 r_g$. The fourth panel of \autoref{fig:timestream_compare} shows that in this vertically limited thermal pressure-dominated zone, 
propagating fluctuations emerge. Moreover, in this particular simulation, waves are also present at smaller radii where the midplane regions are magnetically dominated.  We discuss this in more detail in \autoref{sec:fastmodes} below.  

Finally, \twelve\ is a fully magnetically dominated disk throughout its structure with mean field Maxwell stresses dominating the angular momentum transport. The bottom panel of \autoref{fig:timestream_compare} shows no evidence of propagating temperature waves at any radius, further reinforcing the conclusion that mean-field Maxwell stresses do not excite the same wave patterns seen in simulations with turbulence.  

While most of the waves we find are present in thermal pressure-dominated disks with weak-field MRI turbulence, we now address exactly under what conditions these fluctuations can sometimes occur in magnetically dominated regions. 

\subsection{Fast modes in magnetically dominated disks}
\label{sec:fastmodes}

\begin{figure*}
    \centering
    \includegraphics[width=\textwidth]{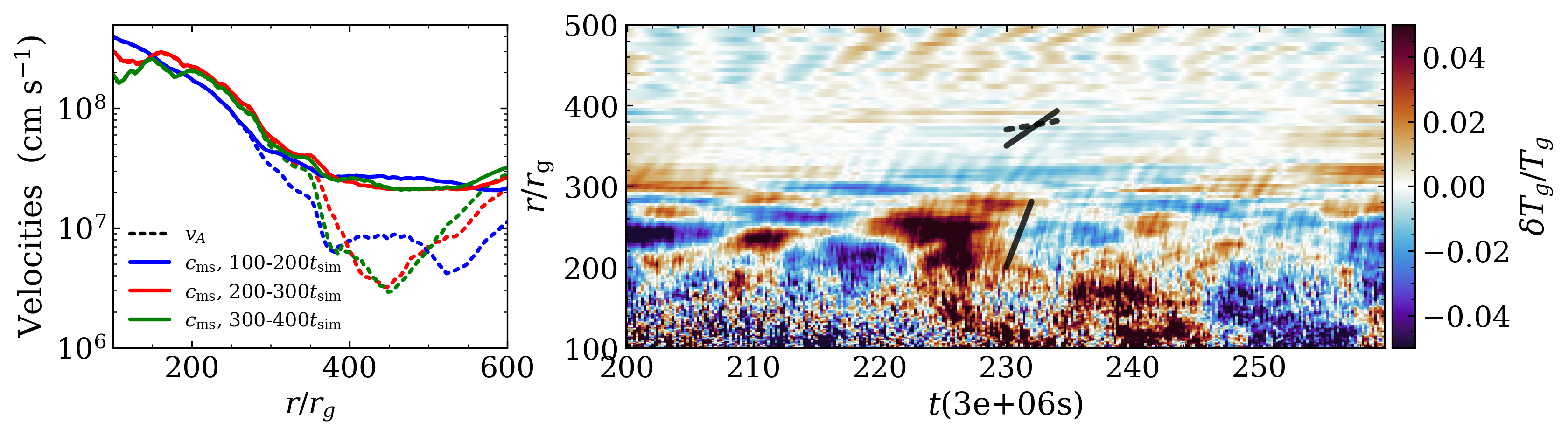}
    \caption{Distinguishing fast magnetosonic modes and thermal sound modes in \sixteenB. (Left) Time- and azimuthally-averaged profiles of  magnetosonic (solid lines) and Alfv\'en velocity (dashed lines)  in the midplane as a function of radius. In the inner, magnetic pressure dominated regions, the Alfv\'en speed matches the magnetosonic speed. Between 350 $r_g$ and 600 $r_{g}$, the midplane is thermal pressure dominated, where the thermal sound speed dominates the magnetosonic speed and thus $v_A\ll c_s$. (Right) Radius-time midplane plot of gas temperature fluctuations in  \sixteenB. Overplotted are the magnetosonic $c_{\rm ms}$ (solid lines) and Alfv\'en speed $v_{A}$ (dashed line) slopes at the corresponding radii. The fluctuations in the inner region propagate at the fast mode speed whereas those in the $r>350r_{g}$ regions propagate at the thermal sound speed.}
    \label{fig:16B_waves}
\end{figure*}

Although \sixteenB\ is magnetically dominated at small radii $(r<350 r_{\rm g}$) in the midplane, \autoref{fig:timestream_compare} shows propagating fluctuations originating in this inner region. 
These particular fluctuations are fast magnetosonic modes as opposed to thermal pressure-dominated sound waves, as shown in \autoref{fig:16B_waves}. The left panel shows the time- and azimuthally-averaged profiles of the magnetosonic speed $c_{\rm ms} = \sqrt{(v_{\rm A}^2 + c_{\rm s}^2)}$ and Alfv\'en speed $v_{\rm A}$ as a function of radius in the midplane. For small radii ($r<350r_g$), magnetic pressure dominates and $c_{\rm ms}\approx v_{\rm A}$ as expected. In contrast, for $r>350 r_{\rm g}$, 
thermal pressure dominates and thus $v_{\rm A}\ll c_{\rm ms}\sim c_{\rm s}$. These slopes for the corresponding radii are plotted on top of the radius-time gas temperature fluctuation plots in the right panel of \autoref{fig:16B_waves}. The solid lines indicate $c_{\rm ms}$ which is the same as $v_{\rm A}$ for smaller radii, and is consistent with the propagation speeds. For larger radii, the propagation speeds  become consistent with $c_{\rm ms}\sim c_{\rm s}$ (solid line) as opposed to $v_{\rm A}$(dashed line).

This case of simulation \sixteenB\/ illustrates that waves that drive thermal temperature fluctuations can exist even in magnetically dominated regions under certain conditions.  As we discuss in detail in \citet{jiang2025}, the magnetically dominated region that exhibits waves in \sixteenB\/ happens to have Maxwell stresses that are dominated by turbulence, as opposed to mean field Maxwell stresses.  In all the other magnetically dominated regions in all our simulations, angular momentum transport is dominated by mean field Maxwell stresses.  Hence the criterion for the existence of waves in all our simulations lies in the level of turbulence in the disk.  In the radiation pressure dominated simulations, namely AGNIron, early \sixteen, \sixteenB\ at large radii, and \eighteenB, weak field MRI turbulence dominates the Maxwell stresses.  And turbulence still dominates the magnetically dominated regions that exhibit waves (now fast magnetosonic waves as opposed to simple thermal pressure dominated sound waves) in \sixteenB.  In the absence of a significant turbulent Maxwell stress, waves of temperature fluctuations are simply not present. Previously, for \eighteenB\ we highlighted the importance of radiation damping, where acoustic waves get damped if the damping timescale is short compared to the wave propagation timescale. Since the waves discussed here in \sixteenB\ are magnetic pressure supported fast waves at radii down to 200$r_{\rm g}$, these should 
be less affected by radiation damping. For magnetosonic waves supported by magnetic pressure, the damping rate from \autoref{eq:fastdamping} decreases by a factor of $\sim P_B/P_{\rm rad}$. Using this expression, we verify that $t_{\rm damp}/t_{\rm wave}\gg1$ for small radii in \sixteenB\ corresponding to the magnetic pressure supported region, where we do observe waves. Thus, fast modes in a magnetic pressure supported region can exist since they are only weakly damped by radiation diffusion. 
A natural consequence of this result is that the photosphere of the observed AGN systems with waves must be magnetic pressure dominated, reflective of fast modes. If this were not the case, strong radiation damping at the photosphere would destroy 
acoustic waves and no fluctuations would be observed.


\section{Discussion}
\label{sec:discussion}

\begin{figure}
    \centering
    \includegraphics[width=\linewidth]{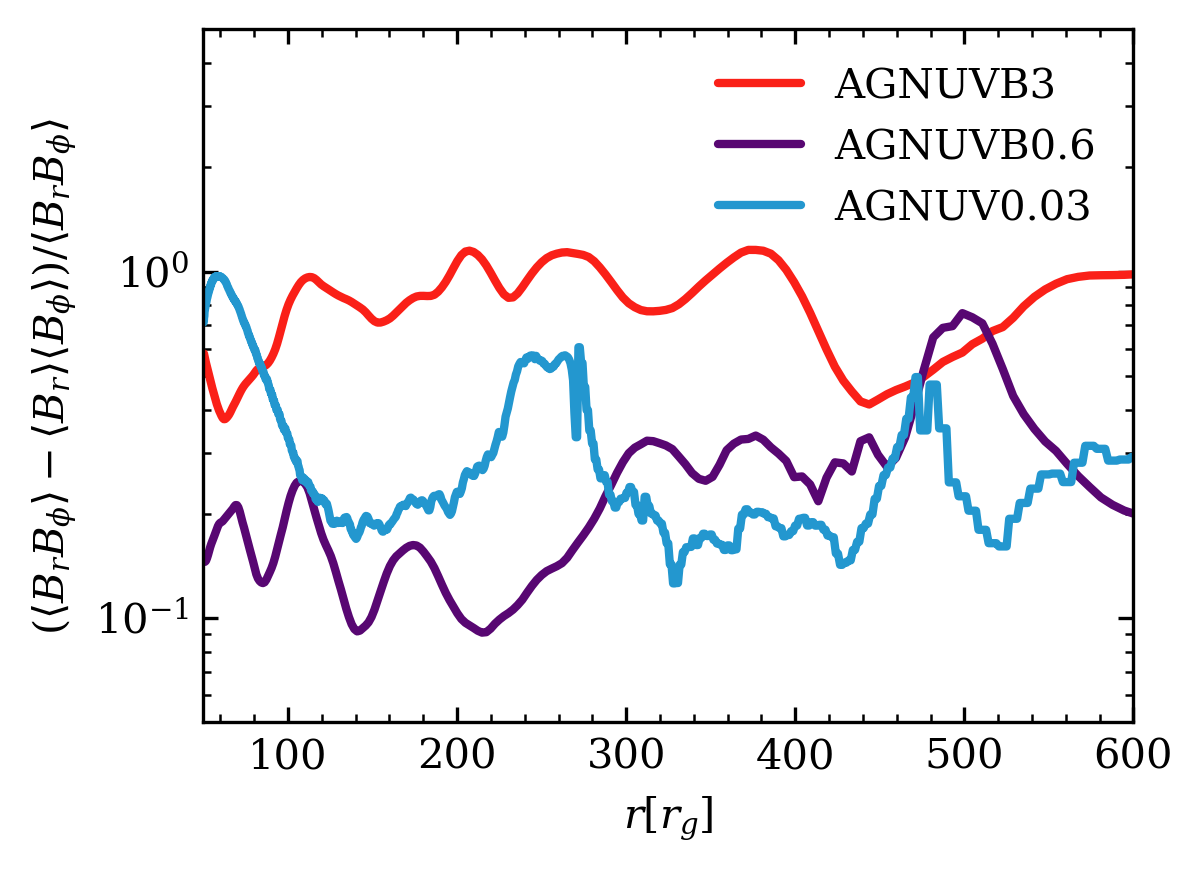}
    \caption{Time- and azimuthally-averaged ratio of turbulent Maxwell stress to the total Maxwell stress in and around the midplane for the various different simulations. A lower ratio implies less turbulence and corresponds to regions of no waves. The exceptions in \twelve\ are the spikes around $r\sim250r_g$ and $470r_g$, which are not turbulence but are instead due to 
    fluctuations in the location of the current sheet around the midplane in this single loop simulation.}
    \label{fig:stress_ratio}
\end{figure}

The propagating fluctuations identified in our simulations closely resemble the variability observed in AGN light curves. \citet{neustadt2022} reported coherent temperature fluctuations, identifying wave-like perturbations that propagate inward and outward at velocities of $\sim 0.01-0.1c$.  Our simulations confirm that such slow-moving fluctuations can naturally arise in both thermal and magnetic pressure-dominated disks as long as the turbulent Maxwell stress is significant, providing a physical explanation for these observations. This is because these wave signatures only arise in the presence 
of turbulence, which is well quantified in our simulations by how significant the turbulent Maxwell stress component is compared to the other stresses. To make this more quantitative, we plot the ratio of turbulent Maxwell stress to total Maxwell stress in these simulations in \autoref{fig:stress_ratio}. Notice that this ratio in \eighteenB\ is usually of order unity indicating a high level of turbulence, which corresponds to having the clearest wave signatures. Similarly in \sixteenB\ we see wave signatures beyond $r\sim 250 r_g$ which is where the ratio increases to $>0.3$. \twelve, which shows no evidence of these waves, has a small ratio except at $r\sim250 r_g$ and $r\sim470r_g$. These spikes however, correspond to fluctuations in the location of the current sheet in this single-loop simulation about the midplane.  Indeed, 
a $\sim2$ orders of magnitude steep jump in the ``turbulent" Maxwell stress occurs at exactly the midplane. We can therefore safely rule out this spike being caused by actual turbulence.

Furthermore, the amplitude and coherence of our simulated fluctuations, with temperature variations of
$2 - 4 \%$, align well with the wave-like signatures in multi-wavelength monitoring campaigns \citep{neustadt2022,stone2023, neustadt2024}.  This agreement suggests that compressible acoustic or fast magnetosonic waves, excited by turbulence, are behind this class of observed continuum variability.  

\citet{heinemann2009-theory} analytically showed, using a WKB approximation on shearing box simulations, that nonaxisymmetric spiral density waves have rms mean velocities that are positively correlated with the level of MRI turbulence in the disk. They confirmed these with numerical shearing box simulations as well \citep{heinemann2009-sim}. While we have only analyzed axisymmetric waves here, it may be that turbulence is capable of exciting such waves as well. Our simulations indicate that in magnetic pressure supported disks, wave-like temperature fluctuation signatures are not excited 
unless the disk has 
a significant turbulent Maxwell stress component regardless of the strength of radiation damping.  Waves 
appear readily in thermal pressure dominated disks 
where weak-field MRI turbulence is always present, provided radiation damping is weak. This is also consistent with findings by \citet{heinemann2009-theory}, because stronger turbulence in the disk implies that
  waves would be more readily observed due to stronger fluctuations. 
\subsection{Observability of waves at the photosphere}

While we find wave signatures of gas temperature fluctuations in the midplane regions of some of our simulations that are remarkably similar in amplitude and time scale to those observed in AGN, we do not find such signatures at the photosphere in any of our simulations.  This is true not only for gas temperature but in fact for any fluid variable, including density and pressure. We describe below three challenges that we face (both physical and numerical in nature) to make these waves observable at the photosphere in our simulations. The former suggest physical constraints on conditions at the photosphere that should be present in AGN disks where these wave patterns have been observed through reverberation mapping.

For thermal pressure-dominated photospheres, radiative damping from photon diffusion will be much faster than the wave period, which will destroy acoustic waves. 
However, when the disk photosphere is magnetic pressure dominated, we have just demonstrated these waves can in principle exist as radiative damping is greatly reduced.  However, in order for these waves to manifest as observable fluctuations in gas temperature, they must be out of LTE with the radiation because diffusion smooths out the radiation temperature fluctuations. The condition for this to be true is given by \autoref{eq:nltecondition} in the Appendix, and we find that it is not satisfied in any of our simulations with the exception of \sixteen, which at late times (when magnetic pressure dominates at the photosphere) does not exhibit waves anyway because mean field Maxwell stresses dominate over turbulence.

Second, our results show that turbulence is necessary in order to excite these waves, and this is mostly present in the midplane regions.  Because the magnetosonic speed increases with altitude, waves near the midplane will be refracted back toward the midplane, regardless of whether they propagate radially inward or outward.  It may therefore be difficult for waves excited in the midplane regions to be able to extend outward to the photosphere.

Third, since we do not observe waves in any fluid variable at the photosphere, even when magnetic pressure dominates, there might be a limitation due to numerical resolution. Indeed, in both simulations \eighteenB\ (\autoref{fig:16B_waves} and movie \href{https://doi.org/10.5281/zenodo.15053915}{online})  and \sixteenB\ (movie \href{https://doi.org/10.5281/zenodo.15498854}{online}), the propagating fluctuations are confined within an angular boundary that closely matches a mesh refinement boundary in those
simulations.  Changing the location of this boundary to be outside the photosphere, or increasing the total resolution, is likely necessary in order for there to be any possibility of manifesting these waves at the photosphere.

\subsection{Stress and Pressure support}
Our suite of magnetic and radiation pressure dominated disks show that whenever Maxwell stress is turbulent, coherent magnetosonic modes are excited independent of whether pressure support is provided by radiation or magnetic fields. 
A highly magnetized disk state, in which large scale mean field Maxwell stresses dominate the angular momentum transport, appears not to excite the thermal fluctuations described in this work. We examined two additional simulations of magnetically dominated disks and found no evidence of the wave-like temperature fluctuations that characterize our results. This suggests that, at least for the AGNs that have been analyzed in these observational studies \citep{neustadt2022, stone2023, neustadt2024}, the disk should either be thermal pressure dominated with weak-field MRI turbulence, or magnetic pressure dominated with a significant turbulent stress component.  How significant the latter should be remains an open question.  Turbulence is the crucial driver of the variability mechanism identified here.  It is interesting that recent cosmological simulations that form magnetic pressure dominated disks \citep{hopkins2024b, hopkins2025} retain a significant turbulent component of the Maxwell stress, and therefore could be expected to drive such temperature fluctuation patterns.
 

Magnetic pressure (and therefore the Alfv\'en speed) is notoriously difficult to observationally constrain in AGN disks.  However, thermal pressure, which depends on temperature, can be observationally constrained.  In a radiation pressure dominated environment, the sound speed depends on density as well as temperature.  If that can also be observationally constrained, then one can distinguish magnetically dominated fast waves from radiation pressure acoustic waves from observation.



\subsection{Caveats and conclusions}
We conclude that the existence of waves in AGN disks hinges on the strength of the turbulent Maxwell stress, independent of whether it is magnetic or radiation pressure dominated. Radiation pressure-dominated disks more readily develop these waves because weak-field MRI drives strong turbulence, but magnetic pressure-dominated disks could also support them as long as the have a significant turbulent stress component. The key requirement, however, is that the photosphere must be magnetic pressure dominated and the gas and radiation must be out of LTE, otherwise radiation damping will smooth out these fluctuations.

While our simulations provide strong evidence for the presence of intrinsic temperature fluctuations in AGN accretion disks, several key questions remain unanswered. Future studies should explore the dependence of these fluctuations on disk parameters such as black hole mass, accretion rate, and magnetic field strength. Furthermore, a parameter survey of global simulations with different Eddington ratios would also help relate observations to simulations. In particular, the observed AGN in \cite{neustadt2022, stone2023, neustadt2024} all have sub-Eddington accretion, with $\lambda_{\rm Edd}\sim 0.01-1$\footnote{except Mrk 142 in \cite{neustadt2022}, which has $\lambda_{\rm Edd}=25$ and shows no evidence of propagating fluctuations.}. Therefore, more simulations in the sub-Eddington regime are imperative, especially since our lowest Eddington ratio simulation \twelve\ does not have significant turbulent Maxwell stresses and therefore shows no propagation patterns. A sub-Eddington disk with significant turbulence would still presumably excite propagating waves.  

Since turbulence is a key feature for these waves to occur, the convergence of the turbulent cascade and its properties with resolution is an important consideration. This has been extensively studied in the context of turbulence driven by weak-field MRI, especially for configurations where the initial condition has no net vertical flux threading the disk \citep{fromang2007, ryan2017}. In these shearing box simulations, it has been found that without explicit viscosity or magnetic diffusivity MRI turbulence remains unconverged \citep{held2022}. In the radiation pressure dominated runs presented in this paper, where turbulent Maxwell stresses dominate, MRI is certainly an important driver of turbulence. 
In the cases where magnetic pressure dominates, it is not the weak field MRI that is driving turbulence, but other processes such as dynamos \citep{squire2024} which might also have problems with numerical convergence. Dedicated studies that examine the impact of numerical resolution on wave amplitudes excited by turbulence should be performed.  

One shortcoming of our simulations is that the fluctuations are not present at altitude near the photosphere. As we discussed, this could be due to strong radiation damping, but it is also possible that this is simply 
due to poor numerical spatial resolution there.  This makes 
it challenging to bridge directly to 
observations. Wave propagation analyses in simulations with better-resolved photospheres 
would help resolve this problem. However, due to the strong radiation damping at the photosphere, we conclude that in AGN systems where these waves are observed the photosphere must be magnetic pressure dominated. This is because only if magnetic pressure dominates could these waves be sustained as fast modes with high radiation damping. If they were radiation pressure dominated, damping would destroy 
these waves and none would be observed.  
An additional condition that must be satisfied in order for magnetically dominated waves to produce observable fluctuations in gas temperature in the presence of radiation damping is that the gas fluctuations must be out of LTE with the radiation.

Overall, our results highlight the need for a more comprehensive framework that integrates both MHD turbulence and radiation pressure effects into AGN disk models. By bridging the gap between simulations and observations, we can develop a deeper understanding of the physical processes that govern AGN variability.

\begin{acknowledgments}
We thank Lunan Sun for some preliminary contributions related to this work.  This work was supported in part by NASA Astrophysics Theory Program grant 80NSSC22K0820.  Resources supporting this work were provided by the NASA High-End Computing (HEC) Program through the NASA Advanced Supercomputing (NAS) Division at Ames Research Center.
\end{acknowledgments}

%

\vspace{5mm}


\software{astropy \citep{astropy},
          numpy \citep{numpy}, matplotlib \citep{matplotlib}, cmasher \citep{cmasher}
          }


\appendix
\section{Radiative Damping of Acoustic and Fast Magnetosonic Waves}

\cite{blaes2003} considered linear perturbations of a static, stratified optically thick medium with a uniform background magnetic field, accounting for diffusive photon transport and possible departures from gas-radiation LTE in the perturbations.  They assumed, however, that the equilibrium state is in LTE so that the gas and radiation effective temperatures are the same.  This appears
to be a valid approximation all the way out to the photosphere for all the simulations presented in this paper, with the exception of \sixteen\/ at late times which does not in any case exhibit coherent waves.
We use their results here to derive diffusive radiative damping rates for magnetosonic waves in a homogeneous medium.

Setting all equilibrium gradient terms to zero in eqs. 40-44, 49, and 53 of \citet{blaes2003}, we obtain the following.  First, the dispersion relation for hydrodynamic sound waves is
\begin{equation}
    \omega^2=k^2({\cal A}c_{\rm g}^2+{\cal B}c_{\rm r}^2+{\cal C})
\end{equation}
and the dispersion relation for magnetosonic waves is (Alfv\'en waves are unaffected by this
thermodynamics)
\begin{equation}
    0=\omega^4-\omega^2k^2v_{\rm A}^2-[\omega^2-({\bf k}\cdot{\bf v}_{\rm A})^2]k^2({\cal A}c_{\rm g}^2+{\cal B}c_{\rm r}^2+{\cal C}).
\end{equation}
Here $\omega$ is the wave angular frequency, ${\bf k}$ is the wave vector, and ${\bf v}_{\rm A}$ is the vector Alfv\'en speed.  For wave propagation perpendicular to the magnetic field (as in the axisymmetric waves
propagating perpendicular to the predominately azimuthal field explored in this study), the fast magnetosonic dispersion relation is simply
\begin{equation}
    \omega^2=k^2(v_{\rm A}^2+{\cal A}c_{\rm g}^2+{\cal B}c_{\rm r}^2+{\cal C}).
    \label{eq:dispfast}
\end{equation}
The quantity ${\cal A}c_{\rm g}^2+{\cal B}c_{\rm r}^2+{\cal C}$ in these dispersion relations is a complex effective squared sound speed given by
\begin{equation}
    {\cal A}c_{\rm g}^2+{\cal B}c_{\rm r}^2+{\cal C}=\frac{\omega^2p(c_{\rm r}^2+c_{\rm g}^2)+\omega\frac{ick^2}{3\kappa_{\rm F}\rho}c_{\rm g}^2p+i\omega_{\rm a}\left\{\omega c_{\rm s}^2[p+4(\gamma-1)E]+\frac{ick^2}{3\kappa_{\rm F}\rho}c_{\rm i}^24E(\gamma-1)\right]}{\omega p\left(\omega+\frac{ick^2}{3\kappa_{\rm F}\rho}\right)+i\omega_{\rm a}\left\{\omega[p+4(\gamma-1)E]+\frac{ick^2}{3\kappa_{\rm F}\rho}4(\gamma-1)E\right\}},
    \label{eq:cs2effective}
\end{equation}
where $p$ is the gas pressure, $E$ is the radiation energy density, $\rho$ is the gas density, and $\kappa_{\rm F}$ is the flux-mean opacity (which we take to be the Rosseland mean opacity in our simulations).  The various speeds in this equation are defined as follows:  $c_{\rm s}^2=\Gamma_1(p+E/3)/\rho$ is the squared adiabatic sound speed for gas and radiation in LTE ($\Gamma_1$ is the first generalized adiabatic exponent for a gas-radiation mixture; \citealt{Chandra1967}), $c_{\rm i}^2=p/\rho$ is the squared isothermal sound speed in the gas, $c_{\rm g}^2=\gamma p/\rho$ is the squared adiabatic sound speed in the gas alone ($\gamma$ is the gas adiabatic exponent), and $c_{\rm r}^2=4E/(9\rho)$ is the squared adiabatic sound speed in the radiation alone.  Thermal coupling between the gas and radiation occurs with an inverse time-scale give by (\citealt{blaes2003}, eq. 30)
\begin{equation}
    \omega_{\rm a}=\left[\kappa_{\rm P}\left(1+\frac{1}{4}\frac{\partial\ln\kappa_J}{\partial\ln T_{\rm r}}\right)+\kappa_{\rm T}\frac{k_{\rm B}T}{m_{\rm e}c^2}\right]\rho c.
\end{equation}
Here $\kappa_{\rm P}$ is the Planck mean opacity, $\kappa_{\rm J}$ is the $J$-mean opacity (which has distinct dependencies on gas temperature $T_{\rm g}$ and radiation temperature $T_{\rm r}$ even if those temperatures are equal in the equilibrium), and $\kappa_{\rm T}$ is the Thomson opacity.  Our simulations do not distinguish Planck and $J$-means, so for our estimates in this paper, we approximate the thermal coupling frequency as
\begin{equation}
    \omega_{\rm a}=\left(\kappa_{\rm P}+\kappa_{\rm T}\frac{k_{\rm B}T}{m_{\rm e}c^2}\right)\rho c.
\end{equation}
The first term in $\omega_{\rm a}$ represents thermal coupling through true absorption and emission processes, while the second term represents thermal coupling through Compton scattering.

Equation (\ref{eq:cs2effective}) captures the essential thermodynamics of these waves.  Photon diffusion acts to smooth out radiation temperature fluctuations if the diffusion frequency $ck^2/(3\kappa_{\rm F}\rho)$ exceeds the wave frequency $\omega$, and whether or not gas temperature fluctuations are also smoothed out depends on the thermal coupling frequency $\omega_{\rm a}$.  For infinite thermal coupling
between the radiation and the gas ($\omega_{\rm a}\rightarrow\infty$), the effective sound speed ranges from the total adiabatic sound speed $c_{\rm s}$ in the gas-radiation mixture at long wavelengths ($k\rightarrow0$, slow diffusion) to the isothermal gas sound speed $c_{\rm i}$ for short wavelengths ($k\rightarrow\infty$, fast diffusion).  In the latter case, all temperature perturbations (and therefore radiation pressure perturbations) are wiped out by diffusion, leaving only an isothermal acoustic response in the gas.  For zero thermal
coupling ($\omega_{\rm a}=0$), the effective sound speed ranges from $(c_{\rm r}^2+c_{\rm g}^2)$ at long wavelengths (slow diffusion - here the gas and radiation undergo separate adiabatic evolution with no constraint on their temperatures being equal) to $c_{\rm g}^2$ (rapid diffusion - here the gas temperature perturbations remain adiabatic as there is no coupling to the radiation).

If we first assume either tight thermal coupling ($\omega_{\rm a}\rightarrow\infty$) or
negligible thermal coupling ($\omega_{\rm a}\rightarrow0$), and negligible gas pressure, the dispersion relation
(\ref{eq:dispfast}) has a solution for long wavelengths given by
\begin{equation}
\omega\simeq
\begin{cases}
    \pm k(v_{\rm A}^2+c_{\rm s}^2)^{1/2}-i\frac{ck^2}{6\kappa_{\rm F}\rho}\left(\frac{c_{\rm s}^2}{v_{\rm A}^2+c_{\rm s}^2}\right)+{\cal O}(k^3), &\text{for $k\rightarrow0$}\\
    \pm kv_{\rm A}-i\frac{2\kappa_{\rm F} E}{3c}+{\cal O}(k^{-1}), &\text{for $k\rightarrow\infty$}
\end{cases}
\label{eq:fastdamping}
\end{equation}
(Note that $c_{\rm s}=c_{\rm r}=(4E/9\rho)^{1/2}$ for $E\gg p$.)
At long wavelengths, the damping rate is reduced by a factor of approximately the magnetic to radiation energy densities for magnetically supported fast waves compared to radiation pressure supported sound waves, allowing magnetically supported fast waves to exist for shorter wavelengths compared to radiation sound waves.  Even if they persist, however, one must ask whether there will be significant temperature perturbations associated with them.

In terms of the density perturbations, the temperature perturbations in the radiation and gas are given by
\begin{equation}
    \frac{\delta T_{\rm r}}{T}=\frac{\omega(\gamma-1)(3p+4E)-i\frac{\omega^2}{\omega_{\rm a}}p}{12(\gamma-1)E\left(\omega+i\frac{ck^2}{3\kappa_{\rm F}\rho}\right)+3\omega p-3p\frac{i\omega}{\omega_{\rm a}}\left(\omega+i\frac{ck^2}{3\kappa_{\rm F}\rho}\right)}\frac{\delta\rho}{\rho},
\end{equation}
\begin{equation}
    \frac{\delta T_{\rm g}}{T}=\frac{-i\frac{\omega}{\omega_{\rm a}}p\left(\omega+i\frac{k^2c}{3\kappa_{\rm F}\rho}\right)+\frac{1}{3}\omega(4E+3p)}{4E\left(\omega+i\frac{k^2c}{3\kappa_{\rm F}\rho}\right)+\frac{\omega}{\gamma-1}p-i\frac{\omega}{\omega_{\rm a}}\frac{p}{\gamma-1}\left(\omega+i\frac{k^2c}{3\kappa_{\rm F}\rho}\right)}\frac{\delta\rho}{\rho}
\end{equation}
respectively.  Note that as $\omega_{\rm a}\rightarrow\infty$, these become
\begin{equation}
    \frac{\delta T_{\rm r}}{T}=\frac{\delta T_{\rm g}}{T}=\frac{\omega(\gamma-1)(3p+4E)}{12(\gamma-1)E\left(\omega+i\frac{k^2c}{3\kappa_{\rm F}\rho}\right)+3\omega p}\frac{\delta\rho}{\rho}
\end{equation}
and both temperature perturbations then become negligible in the rapid diffusion regime.
For $\omega_{\rm a}\rightarrow0$, we instead obtain
\begin{equation}
    \frac{\delta T_{\rm r}}{T}=\frac{\omega}{3\left(\omega+i\frac{ck^2}{3\kappa_{\rm F}\rho}\right)}\frac{\delta\rho}{\rho},
\end{equation}
and
\begin{equation}
    \frac{\delta T_{\rm g}}{T}=(\gamma-1)\frac{\delta\rho}{\rho},
    \label{eq:Tgasadiabatic}
\end{equation}
where the latter just corresponds to the usual adiabatic relation for the gas alone.  The former corresponds to the usual adiabatic relation for the radiation alone for long wavelengths, but $\delta T_{\rm r}/T\rightarrow0$ for short wavelengths because of rapid
radiative diffusion.

The low absorption frequency limit (\ref{eq:Tgasadiabatic}) is achievable provided
\begin{equation}
    \omega_{\rm a}\left(1+\frac{4E}{3p}\right)\ll\left|\omega+i\frac{ck^2}{3\kappa_{\rm F}\rho}\right|
    \label{eq:nltecondition}
\end{equation}
regardless of the diffusion regime (rapid or slow).  Inequality (\ref{eq:nltecondition}) must be satisfied in order for significant gas
temperature fluctuations to exist near the photosphere.


\bibliography{refs_ish}{}
\bibliographystyle{aasjournal}



\end{CJK*}

\end{document}